\documentclass[a4paper,10pt]{article}
\usepackage{amssymb} 
\usepackage{latexsym}
\def\Doublehat#1{\skew3\widehat{\widehat{#1}}}
\begin{document}
\begin{center}
\LARGE { A Duality for Yang-Mills Moduli Spaces on Noncommutative Manifolds }\\
\vspace{2cm}
\Large { Hiroshi TAKAI} \\
\vspace{5mm}
\Large { Department of Mathematics, \\
Tokyo Metropolitan University} \\

\end{center}
\vspace{2cm}

\begin{center}
\Large  Abstract 
\end{center}
\large  \quad  Studied are the moduli spaces of Yang-Mills connections on 
finitely generated projective modules associated with noncommutative flows. 
It is actually shown that they are homeomorphic to those on dual modules 
associated with dual noncommutative flows. As a corollary, the result is also 
affirmative to the case of noncommutative multiflows. 
 As an important application, computed are the the moduli spaces of the 
instanton bundles over noncommutative Euclidean 4-spaces with respect to the 
canonical action of space translations without using the ADHM-construction.

\newpage
\Large{\S1.~Introduction} \large \quad Among miscellaneous topics in super 
string theory or M-theory, one of their most important problems is concerned 
with the compactification of fields, which means that either 10 or 11 
dimensional field theory would be reduced to 4 dimensional one by compactifying either 6 or 7 dimensional space time respectively. For instance, an 11 
dimensional M-theory has a circle compactification to deduce a IIA-type super 
string theory, which describes a nonchiral field theory of closed strings due 
to BFSS ([9]). Moreover, this theory has also one more circle compactification
to deduce a IIB-type superstring theory, which describes a chiral field theory 
of closed strings via the so-called T-transformations ([10]). Recently, Connes, Douglas and Schwarz [2] have shown that the field theory to such a 2-torus 
compactification cited above has a complete solution by taking the moduli spaces of Yang-Mills connections of appropriate modules for the gauge action of the 
2-torus on either commutative or noncommutative 2-torus. Actually, Connes and 
Rieffel [3] have proved that the latter Yang-Mills moduli space is homeomorphic  to the 2-torus. From this point of view, the problem of finding the Yang-Mills  moduli space for a given smooth noncommutative dynamical system is a quite 
important one to determine the unified 4 dimensional field theory having the 
unique compactification. \\
In this paper, we present a certain duality of Yang-Mills moduli spaces for 
noncommutative flows. More precisely, we show that the Yang-Mills moduli spaces for smooth noncommutative flows are homeomorphic to those for the associated 
dual flows. This could be interpreted as no physical data is changed under 
dimension reduction of space time. The method itself is also applicable to 
noncommutative multi flows in principle. As an important application, we 
determine topologically the moduli spaces of the instanton bundles over 
noncommutative Euclidean 4-spaces with respect the canonical action of the space translations without using the ADHM-construction (cf:[5],[7],[8]). \\
 
\Large{\S2.~Noncommutative Yang-Mills Theory} \large ~In this section, we review  the noncommutative Yang-Mills theory due to Connes-Rieffel[3]. Let $(A,G,\alpha)$ be a C$^*$-dynamical system, $A^\infty$ the set of all smooth elements of 
$A$ under $\alpha$ and $\alpha^\infty$ the restriction of $\alpha$ 
to $A^\infty$ where $G$ is a connected Lie group. Then the system 
$(A^\infty,G,\alpha^\infty)$ becomes a noncommutative smooth dynamical system. 
In what follows, we only treat such a dynamical system, so that we notationally write it by $(A,G,\alpha)$. Let $\delta$ be the differentiation map of $\alpha$. Then it is a Lie homomorphism from the Lie algebra $\mathcal{G}$ of $G$ to the Lie algebra $Der(A)$ of all $^*$-derivations of $A$. Let $\Xi$ be a finitely 
generated projective right $A$-module. Then it has a Hermitian structure 
$< \cdot \mid \cdot >_A$ with the property that
\[< \xi \mid \eta >^*_A = < \eta \mid \xi >_A,~< \xi \mid {\eta}a >_A = < \xi \mid \eta >_A a   \]
\noindent
$(\xi,\eta \in \Xi,a \in A)$. Now we can define a noncommutative version of 
connections on vector bundles over manifolds in the following fashion: 
Let $\nabla$ be a linear map from $\Xi$ to $\Xi \otimes \mathcal{G}^*$. Then it is called a connection of $\Xi$ if it satisfies 
\[ \nabla_X({\xi}a) = \nabla_X(\xi)a + {\xi}\delta_X(a)  \]
\noindent
$(\xi \in \Xi,a \in A,X \in \mathcal{G})$. Moreover, a connection $\nabla$ is 
said to be compatible with respect to $< \cdot \mid \cdot >_A$ (or compatible) 
if it satisfies
\[ \delta_X(< \xi \mid \eta >) = < \nabla_X(\xi) \mid \eta > 
                               + < \xi \mid \nabla_X(\eta) >  \]
\noindent
$(\xi \in \Xi,a \in A,X \in \mathcal{G})$. We denote by $CC(\Xi)$ the set of all compatible connections of $\Xi$. Then it is nonempty because it contains the 
so-called Grassmann connection $\nabla^0$, which is defined as follows: 
By assumption, $\Xi = P(A^n)$ for some $n \geq 1$ and a projection 
$P \in M_n(\mathcal{M}(A))$ where $\mathcal{M}(A)$ is the multiplier algebra of  $A$. Then $\nabla^0 = P[\delta^n]$ becomes a compatible connection of $\Xi$, 
where $\delta^n$ is the differentiation map of the action $\alpha^n = \alpha \otimes id_n$ on $M_n(A)$. Now for any $\nabla \in CC(\Xi)$, there exists an 
element $\Omega_X \in \mathrm{End}_A(\Xi)$ such that
\[  \nabla_X = \nabla^0_X + \Omega_X   \]
\noindent
$(X \in \mathcal{G})$, where $E=\mathrm{End}_A(\Xi)$ is the set of all 
$A$-endomorphisms of $\Xi$. Since $\nabla$ and $\nabla^0$ are compatible, then 
$\Omega_X ~(X \in \mathcal{G})$ are all skew-adjoint. 
Given a $\nabla \in CC(\Xi)$, there exists a skew adjoint $E$-valued 2-form 
$\Theta_{\nabla}$ of $G$ such as 
\[ \Theta_{\nabla}(X,Y) = \nabla_X{\nabla_Y}-\nabla_Y{\nabla_X}-\nabla_{[X,Y]}  \]
\noindent
$(X,Y \in \mathcal{G})$. It is called the curvature of $\nabla$ associated with $(A,G,\alpha)$. Then $\nabla$ is said to be flat if there exists a 2-form 
$\omega$ of $G$ such that 
\[  \Theta_{\nabla}(X,Y) = \omega(X,Y)~Id_{E}  \]
\noindent
$(X,Y \in \mathcal{G})$. We now assume an existance of a continuous $\alpha$-
invariant faithful trace $\tau$ on $A$ as anoncommutative version of
integrability of manifolds. Then there also exists a continuous faithful trace 
$\tilde{\tau}$ on $E$ such that 
\[  \tilde{\tau}(< \xi \mid \eta >_{E}) =\tau(< \eta \mid \xi >_A)  \]
\noindent
$(\xi,\eta \in \Xi)$, where
\[  < \xi \mid \eta >_{E}(\zeta) = \xi < \eta \mid \zeta >_A  \]
\noindent
$(\xi,\eta,\zeta \in \Xi)$. In fact, it is well defined because of the 
assumption of $\Xi$. Using $\tilde{\tau}$, we define a noncommutative version of the Yang-Mills functional on manifolds as follows:
\[  \mathrm{YM}(\nabla) = - \tilde{\tau}(\{{\Theta}_{\nabla}\}^2)  \]
where
\[  \{\Theta_{\nabla}\}^2 = \sum_{i<j}~\Theta_{\nabla}(X_i \wedge X_j)^2 \in E  \]
\noindent
for an orthonormal basis $\{X_i\}_i$ of $\mathcal{G}$ with respect to the 
Killing form. Since $\Theta_{\nabla}(X_i \wedge X_j)$ are all skew adjoint, 
then $\{\Theta_{\nabla}\}^2$ has negative spectra only. 
Therefore $\rm{YM}(\nabla) \geq 0$ for all $\nabla \in CC(\Xi)$. Moreover,
it is independent of the choice of a hermitian structure $<\cdot \mid \cdot>_A$
on $\Xi$. Now let $U(E)$ be the set of all unitaries of $E$. 
It is called the gauge group of $\Xi$. For any $u \in U(E)$, we define the gauge transformation $\gamma_u$ on $CC(\Xi)$ by
\[  (\gamma_u(\nabla))_X(\xi) = u(\nabla_X)(u^*\xi)   \]
\noindent
$(u \in U(E),X \in \mathcal{G},\xi \in$ $\Xi$). Then $\gamma$ calls the gauge 
action of $U(E)$ on $CC(\Xi)$. The Yang-Mills functional $\mathrm{YM}$ is 
$\gamma$-invariant, namely
\[  \mathrm{YM}(\gamma_u(\nabla)) = \mathrm{YM}(\nabla)  \]
\noindent
$(u \in U(E),\nabla \in CC(\Xi))$. We then consider the first variational 
problem of $\mathrm{YM}$, namely find a $\nabla \in CC(\Xi)$ such that 
\[  \frac{d}{dt}(\mathrm{YM}(\nabla_t))\Bigm{\vert}_{t=0} = 0   \]
\noindent
for any smooth path $\nabla_t \in CC(\Xi)~(\vert t \vert<\epsilon)$ with 
$\nabla_0=\nabla$, which is called a Yang-Mills connection of $\Xi$ with 
respect to the system $(A,G,\alpha,\tau)$. Let $MC(\Xi)$ be the set of all Yang-Mills connections of $\Xi$ with respect to $(A,G,\alpha,\tau)$. Then the orbit 
space $\mathcal{M}^{(A,G,\alpha,\tau)}(\Xi)$ of $MC(\Xi)$ by the gauge action 
$\gamma$ of $U(E)$ is called the moduli space of the Yang-Mills connections of 
$\Xi$ with respect to the system $(A,G,\alpha,\tau)$. 
We then state the following theorem due to Connes-Rieffel[3] which is quite 
powerful to construct a Yang-Mills connection: \\

Theorem 2.1([3]) \quad Let $(A,G,\alpha)$ be a $C^{\infty}$-dynamical system and $\tau$ be a faithful $\alpha$-invariant continuous trace on $A$. Let $\Xi$ be a finitely generated projective right $A$-module. If G is an abelian connected 
Lie group, then $\nabla \in MC(\Xi)$ if and only if it is flat for any $\nabla 
\in CC(\Xi)$. \\

\Large{\S3.~Dual Yang-Mills Moduli spaces }  \large \quad 
In this section, we only take Frechet flows (or multi-flows) as a special case 
of C$^{\infty}$-dynamical systems. According to Elliott-Natsume-Nest[4], 
let $(A,\mathbb{R},\alpha)$ be a Frechet $^*$-flow in the sence that 
\begin{enumerate}
  \item[(1)]  $(A,\{\| \cdot \|_n\}_{n \geq 1})$ is a Frechet $^*$-algebra
  			  (which is dense in a C*-algebra),
  \item[(2)]  $t \mapsto \alpha_t(a)$ is $C^{\infty}$-class with respect to
              $\| \cdot \|_n~ (n \geq 1)~,$ 
  \item[(3)]  For any $m,k\geq 1$, there exist $n,j\geq 1$ and $C>0$ such that
\[ \Bigm \| \frac{d^k}{dt^k} \alpha_t(a) \Bigm \|_m  < 
      C(1+t^2)^{j/2}  \|a\|_n \qquad (a \in A,~t \in \mathbb{R})  \]
\end{enumerate}
In what follows, we state Frechet $^*$-flows by F$^*$-flows. Typical are the 
following three examples as F$^*$-flows: \\

Examples 3.1 \quad Let $\mathcal{S}(\mathbb{R})$ be the abelian F$^*$-algabra of all complex valued rapidly decreasing smooth functions on $\mathbb{R}$ and 
$\lambda$ the shift action of $\mathbb{R}$ on $\mathcal{S}(\mathbb{R})$. Then 
the triplet $(\cal{S}(\mathbb{R}),\mathbb{R},\lambda)$ is a F$^*$-flow. \\

Example 3.2 \quad Let $\mathcal{K}^\infty(\mathbb{R})$ be the F$^*$-algebra 
consisting of all compact operators on $L^2(\mathbb{R})$ with their integral 
kernels in $\mathcal{S}(\mathbb{R}^2)$, and $Ad(\lambda)$ the adjoint action of $\mathbb{R}$ on $\mathcal{K}^\infty(\mathbb{R})$. Then the triplet $(\mathcal{K}^\infty(\mathbb{R}),\mathbb{R}$,$Ad(\lambda)$) is a F$^*$-flow. \\

Example 3.3 \quad Let $\mathbb{R}_{\theta}^2$ be the F$^*$-algebra 
$\mathcal{S}$($\mathbb{R}^2)$ with Moyal product $\star_{\theta}~(\theta \in \mathbb{R})$, and $\widehat{\theta}$ the dual action of the canonical action $\theta$ on $\mathcal{S}(\mathbb{R})$. Then the triplet $(\mathbb{R}_{\theta}^2,\mathbb{R},\widehat{\theta})$ is a F$^*$-flow. \\

Now let $(A,\mathbb{R},\alpha)$ be a F$^*$-flow with a continuous $\alpha$-
invariant faithful trace $\tau$, and let $\mathcal{S}(\mathbb{R},A)$ be the 
F$^*$-algebra consisting of all $A$-valued rapidly decreasing smooth functions 
on $\mathbb{R}$ with its seminorms $\|\cdot\|_{m,n}$ given by
\[  \| x \|_{m,n} = \mathrm{sup}_{t \in \mathbb{R}}(1+t^2)^{m/2} 
		\Bigm \| \frac{d^n}{dt^n} x(t) \Bigm \|_m    \]
\noindent
($x \in \mathcal{S}(\mathbb{R},A)$). Moreover it has the following product and 
involution:
 \[   (1)~ (x *_\alpha y)(t) = 
		 \int_\mathbb{R} x(s)\alpha_s(y(t-s))ds , 
	  (2)~ x^*(t) = \alpha_t(x(-t))^*   \]
\noindent	    
($x,y \in \mathcal{S}(\mathbb{R},A)$). Then we call $\mathcal{S}(\mathbb{R},A)$ the F$^*$-crossed product of $A$ by the action $\alpha$ of $\mathbb{R}$, which 
is written by $A\rtimes_\alpha \mathbb{R}$. In fact, the three examples cited 
above are isomorphic to $\mathbb{C}\rtimes_\iota \mathbb{R}$, 
$\mathcal{S}(\mathbb{R})\rtimes_\lambda \mathbb{R}$ and 
$\mathcal{S}(\mathbb{R})\rtimes_\theta \mathbb{R}$ respectively. Then we define two actions $\widehat{\alpha}, \widetilde{\alpha}$ of $\mathbb{R}$ on 
$A \rtimes_\alpha \mathbb{R}$ given by 
\[\widehat{\alpha}_s(x)(t)=e^{2{\pi}ist}x(t),~
  \widetilde{\alpha}_s(x)(t)=\alpha_s(x(t)) \quad (i=\sqrt{-1})  \]
\noindent
$(x \in A\rtimes_\alpha \mathbb{R},~s,t \in \mathbb{R})$. The triplets $(A\rtimes_\alpha \mathbb{R},\mathbb{R},\widehat{\alpha})$ and $(A\rtimes_\alpha \mathbb{R},\mathbb{R},\widetilde{\alpha})$ become F$^*$-flows. The former is called to 
be the dual F$^*$-flow of $(A,\mathbb{R},\alpha)$. Then the same duality holds as in the case of $C^*$-crossed products in the following: \\

Theorem 3.4([4]) \quad Given a F$^*$-flow $(A,\mathbb{R},\alpha)$, its double 
dual F$^*$-flow $(A \rtimes_\alpha \mathbb{R} \rtimes_{\widehat{\alpha}} 
\mathbb{R},\mathbb{R},\Doublehat{\alpha})$ is isomorphic to the F$^*$-flow 
$(A \otimes \mathcal{K}^\infty(\mathbb{R}),\mathbb{R},\alpha \otimes Ad(\lambda))$. \\
\noindent
In fact, the equivariant isomorphism $\Psi_{\alpha}^0: 
A \rtimes_\alpha \mathbb{R} \rtimes_{\widehat{\alpha}} \mathbb{R}
\longmapsto A \otimes \mathcal{K}^\infty(\mathbb{R})$ is given by
\[ \Psi_{\alpha}^0(x)(t,s) = 
			\int_{\mathbb{R}} e^{2{\pi}irs}\alpha_{-t}(x(t-s,r))~dr  \]
\noindent
$(x \in A \rtimes_{\alpha} \mathbb{R} \rtimes_{\widehat{\alpha}}\mathbb{R},~t,s  \in \mathbb{R})$. Then the inverse isomorphism $(\Psi_{\alpha}^0)^{-1}$ of 
$\Psi_{\alpha}^0$ is given by
\[ (\Psi_{\alpha}^0)^{-1}(x)(t,s) = 
		\int_{\mathbb{R}} e^{2{\pi}i(t-r)s}\alpha_r(x(r,r-t))~dr   \]
\noindent
$(x \in A \otimes \mathcal{K}^\infty(\mathbb{R}),~t,s \in \mathbb{R})$. 
Now let $(A \rtimes_\alpha \mathbb{R},\mathbb{R},\widehat{\alpha})$ be the dual F$^*$-flow of $(A,\mathbb{R},\alpha)$. If there exists a continuous faithful 
$\alpha$-invariant trace $\tau$ on $A$, then so does it for the F$^*$-flow 
$(\widehat{A},\mathbb{R},\widehat{\alpha})$ given by
\[ 	 \widehat{\tau}(x) ~=~ \tau(x(0))   \qquad (x \in \widehat{A})  \]
\noindent
where $\widehat{A} = A \rtimes_\alpha \mathbb{R}$. Then $\widehat{\tau}$ is 
called the dual trace of $\tau$. Then we consider the Yang-Mills moduli spaces 
for such dual systems. Namely, let $\Xi$ be a finitely generated projective 
right $A$-module and $\widehat{\Xi} = \Xi \otimes_A \widehat{A}$. Then it 
becomes a finitely generated projective right $\widehat{A}$-module. Indeed, the action of $\widehat{A}$ on $\widehat{\Xi}$ is given by 
\[	({\xi}x)(t)=\int_\mathbb{R} \xi(s)\alpha_s(x(t-s))~ds	\]
\noindent
$(\xi \in \widehat{\Xi},x \in \widehat{A})$. On the other hand, the action 
$\widetilde{\alpha}$ is implimented by an unitary multiplier flow on 
$\widehat{A}$, namely there exists a strictly continuous unitary flow $\widetilde{u}$ of the multiplier algebra $\mathcal{\widehat{A}}$ of $\widehat{A}$ 
such that $\widetilde{\alpha}_t=Ad(\widetilde{u}_t)$ on $\widehat{A}$. 
Then it follows that $\widehat{\tau}$ is $\widetilde{\alpha}$-invariant. 
Since the action $\widehat{\alpha}$ commutes with $\widetilde{\alpha}$, we can 
define the action $\overline{\alpha}$ of $\mathbb{R}$ on $\widehat{A}$ by 
$\widehat{\alpha} \circ \widetilde{\alpha}$ which makes $\widehat{\tau}$ 
invariant. Then we propose anothet Yang-Mills moduli space 
$\mathcal{M}^{(\widehat{A},\mathbb{R},\overline{\alpha},\widehat{\tau})}(\widehat{\Xi})$ of $\widehat{\Xi}$ with respect to the F$^*$-flow $(\widehat{A},\mathbb{R},\overline{\alpha})$ and the dual trace $\widehat{\tau}$, 
which is called the dual Yang-Mills moduli space of $\mathcal{M}^{(A,\mathbb{R},\alpha,\tau)}(\Xi)$. \\

\Large{\S4.~Main result} \large \quad In this section, we prove the following 
theorem, which means physically that in quantum field theory, all physical data  are invariant under dimension reduction: \\

Theorem 4.1 [Duality] \quad Let $(A,\mathbb{R},\alpha)$ be a F$^*$-flow 
with a continuous $\alpha$-invariant faithful trace $\tau$ and let $\Xi$ be a 
finitely generated projective right $A$-module. Then there exist a F$^*$-flow 
$(\widehat{A},\mathbb{R},\overline{\alpha})$ with a dual trace 
$\widehat{\tau}$ of $\tau$ and a finitely generated projective right 
$\widehat{A}$-module $\widehat{\Xi}$ whose Yang-Mills moduli space 
${\mathcal{M}}^{(\widehat{A},\mathbb{R},\overline{\alpha},\widehat{\tau})}
(\widehat{\Xi})$ is homeomorphic to 
$\mathcal{M}^{(A,\mathbb{R},\alpha,\tau)}(\Xi)$~. \\

\noindent
Applying Theorems 3.4 and 4.1, we have the following corollary: \\

Corollary 4.2 [Dimension Reduction] \quad  
Let $(\widehat{A},\mathbb{R},\overline{\alpha})$ be the F$^*$-flow cited in Theorem 4.1 and $\beta$ a smooth flow on $\widehat{A}$ commuting with 
$\overline{\alpha}$. Suppose there exists a continuous faithful $\beta$-
invariant trace $\tau$, then given a finitely generated projective right $\beta$-module $\Xi$, there exists a F$^*$-flow $(A,\mathbb{R},\beta_A)$ 
with a continuous faithful $\beta_A$-invariant trace $\tau_A$ and a finitely generated projective $A$-module $\Xi_A$ such that 
$\mathcal{M}^{(\widehat{A},\mathbb{R},\beta,\tau)}(\Xi)$ is homeomorphic to 
$\mathcal{M}^{(A,\mathbb{R},\beta_A,\tau_A)}(\Xi_A)~.$ \\

Proof of Theorem 4.1: \quad By the assumption of $\Xi$, there exist a natural 
number $n$ and a projection $P \in M_n(\mathcal{M}(A))$ such that 
$\Xi = P(A^n)$, where $\mathcal{M}(A)$ is the multiplier algebra of $A$. 
Let us take a Hermitian structure $<\cdot \mid \cdot >_A$ on $\Xi$ by
\[ < \xi \mid \eta >_A  ~=~ 
			\sum^n_{j=1} ~ {\xi}_j^* ~{\eta}_j    \]
\noindent
$(\xi,\eta \in \Xi)$. 
Then if $\nabla^0$ is the Grassmann connection of $\Xi$, then it belongs to 
$CC(\Xi)$. Moreover it follows from Theorem 2.1 that $\nabla^0 \in MC(\Xi)$. 
Now for any $\nabla \in MC(\Xi)$ and $X \in Lie(\mathbb{R})$, there exists a 
skew adjoint element $\Omega_X \in E$ such that 
\[ \nabla_X = \nabla^0_X  + \Omega_X ~. \]  
As $\Xi = P(A^n)$, it follows that 
$\widehat{\Xi} = \widehat{P}(\widehat{A}^n)$ 
where 
\[ \widehat{P}=P \otimes I_{\widehat{A}} \in M_n(\mathcal{M}(\widehat{A}))~.\]
\noindent
Then we know that
\[\mathrm{End}_{\widehat{A}}(\widehat{\Xi})=\widehat{P}M_n(\widehat{A})\widehat{P}~, \]
\noindent
which is denoted by $\widehat{E}$. 
From now on, we want to define a mapping from $\mathcal{M}^{(A,\mathbb{R},\alpha,\tau)}(\Xi)$ into $\mathcal{M}^{(\widehat{A},\mathbb{R},\overline{\alpha},\widehat{\tau})}(\widehat{\Xi})$ in the following way: 
\quad Since $E=\mathrm{End}_{A}(\Xi)$ is no longer $\alpha^n$-invariant in general, it follows using the same idea in Connes[1] that there exists a F$^*$-flow 
$(M_n(A),\mathbb{R},\beta)$ with the property that 
\begin{enumerate}
\item [(1)] \quad $\beta_t(Pa)=P\beta_t(a)~(a \in M_n(A),~t \in \mathbb{R})$,
\item [(2)] \quad $(M_n(A),\mathbb{R},\beta)$ is outer equivalent to 
$(M_n(A),\mathbb{R},\alpha^n)$. 
\end{enumerate} 
By [1], let $\iota_u$ be the equivariant isomorphism from the F$^*$-system 
$(M_n(A) \rtimes_{\alpha^n} \mathbb{R},\widehat{\alpha^n})$ onto $(M_n(A) 
\rtimes_{\beta} \mathbb{R}, \widehat{\beta})$ such that
\[\iota_u \circ \widetilde{\alpha^n} \circ \iota_u^{-1}=\widetilde{\beta}~.\]
\noindent
Then we have the following lemma which would be applied later: \\

Lemma 4.3([1])  \quad The next two statements holds:
\begin{enumerate}
  \item[(1)] There is an equivariant isomorphism $\widetilde{\iota_u}$ from \\
   $(M_n(A) \rtimes_{\alpha^n} \rtimes_{\widehat{\alpha^n}} \mathbb{R},\mathbb{R},\Doublehat{\alpha^n})$ onto 
$(M_n(A) \rtimes_{\beta} \mathbb{R} \rtimes_{\widehat{\beta}},\mathbb{R},\Doublehat{\beta})$.
  \item[(2)]  There exists a unitary multiplier $W$ of 
$M_n(A) \otimes \mathcal{K}^\infty(\mathbb{R})$ such that 
\[ Ad(W) \circ \Psi_{\beta}^0 \circ \widetilde{\iota_u} = \Psi_{\alpha^n}^0 ,\]
\end{enumerate} 				
where $\Psi_{\{ \cdot \}}^0$ are the equivariant isomorphisms as in Theorem 3.4 associated with $\{ \cdot \}$,and 
\[\widetilde{\iota_u}(a)(s)=\iota_u\{a(s)\} \]
\noindent
for all 
\[a \in \mathcal{S}(\mathbb{R},M_n(A) \rtimes_{\alpha^n} \mathbb{R}) ~.\]
\noindent
Since $M_n(\Xi)=(P \otimes I_n)([M_n(A)]^n)$, it is a finitely generated 
projective $M_n(A)$-module. Let $d\beta$ be the infinitesimal generator of 
$\beta$. Now for any $\nabla \in \mathrm{MC}^{(M_n(A),\mathbb{R},\beta,\tau^n)}(M_n(\Xi))$, there exists a skew adjoint element $\Omega \in E_n = \mathrm{End}_{M_n(A)}(M_n(\Xi))$ such that
\[ \nabla = (P \otimes I_n)d{\beta}^n + \Omega . \]
Let $\widehat{E_n}=\widehat{\mathrm{End}}_{M_n(A)}(M_n(\Xi))$ be the set of all $E_n$-valued rapidly decreasing smooth functions on $\mathbb{R}$. Then it 
becomes a F$^*$-algebra with respect to the $\beta^n$-twisted convolution 
product. By definition, we see that 
\[ \widehat{E_n}=\mathrm{End}_{\widehat{M_n(A)}}(\widehat{M_n(\Xi)}) , \]
\noindent
where 
\[\widehat{M_n(A)}=M_n(A) \rtimes_{\beta} \mathbb{R}~,~
\widehat{M_n(\Xi)}=M_n(\Xi) \otimes_{M_n(A)} M_n(\widehat{A})~.\]
\noindent
Let us define the element ${\widehat{\Omega}} \in \widehat{E_n}$ by
\[ {\widehat{\Omega}}(\xi)(t)=\Omega\{\xi(t)\} \]
\noindent
$(\xi \in \widehat{M_n(\Xi)},t \in \mathbb{R})$  In fact, we check that
\[{\widehat{\Omega}}({\xi}a)(t)= \Omega\{({\xi}a)(t)\}
=\int_{\mathbb{R}}~\Omega\{(\xi)(s)\beta_s(a(t-s))\}~ds  \]
\[~~~~~~~~~~~~~~~~~~~~~~~~~~~
=\int_{\mathbb{R}}~\Omega\{\xi(s)\}\beta_s(a(t-s))~ds \]
\[~~~~~~~~~~~~~~~~~~~~~~~~~~~
=\int_{\mathbb{R}}~\widehat{\Omega}(\xi)(s)\beta_s(a(t-s))~ds  \]
\noindent
$~~~~~~~~~~~~~~~~~~~~~~~~~~~~~~~~~~~~~=({\widehat{\Omega}}({\xi}a)(t)$  \\
\noindent
$(\xi \in \widehat{M_n(\Xi)},a \in \widehat{M_n(A)})$. 
As $\beta^n$ is used as the restriction of the natural extension of $\beta$ of 
$M_{n^2}(A))$ to $E_n$, then it follows from the definition that 
\[ \widehat{E_n}=E_n \rtimes_{\beta^n} \mathbb{R} .\] 
Then we obtain that 
\[{\widehat{\Omega}} \in E_n \rtimes_{\beta^n} \mathbb{R}. \]
\noindent
We then have the following lemma:\\

Lemma 4.4 \qquad $\widehat{\Omega} \in \widehat{E_n}$ is skew adjoint.\\

Proof. \quad Since $\Omega \in E_n$ is skew adjoint, we compute that\\

$<\widehat{\Omega}(\xi \otimes f) \mid \eta \otimes g>_{\widehat{M_n(A)}}$
\[=\displaystyle \sum_{j=1}^{n} \widehat{\Omega}(\xi \otimes f)_j^*(\eta \otimes g)_j \].
\noindent
where $\widehat{\Omega}(\xi \otimes f)_j,(\eta \otimes g)_j \in \widehat{M_n(A)}$. Then it is easy to check that
\[ \widehat{\Omega}(\xi \otimes f)_j=\Omega(\xi)_j \otimes f \],
\noindent
using which we deduce that\\

$<\widehat{\Omega}(\xi \otimes f) \mid \eta \otimes g>_{\widehat{M_n(A)}}(t)$
\[=\displaystyle \sum_{j=1}^{n}{\int}_{\mathbb{R}}\beta_s(\Omega(\xi)_j^*\eta_j)\overline{f(-s)}g(t-s)~ds \]
\[~~~~~~~~={\int}_{\mathbb{R}}\beta_s\{<\Omega(\xi) \mid \eta>_{M_n(A)}\}\overline{f(-s)}g(t-s)~ds \]
\noindent
As $\Omega$ is skew adjoint, it follows that
\[<\Omega(\xi) \mid \eta>_{M_n(A)}=-<\xi \mid \Omega(\eta)>_{M_n(A)}\]. 
\noindent
Then we obtain that
\[<\widehat{\Omega}(\xi \otimes f) \mid \eta \otimes g>_{\widehat{M_n(A)}}(t)
=-<\xi \otimes f \mid \widehat{\Omega}(\eta \otimes g)>_{\widehat{M_n(A)}}(t)\]
\noindent
$(\xi,\eta \in M_n(\Xi),~f,g \in \cal{S}(\mathbb{R}))$. This implies the 
conclusion.  ~ Q.E.D. \\

Let $d\widehat{\beta^n}$ and $d\widetilde{\beta^n}$ be the infinitesimal generators of the dual action $\widehat{\beta^n}$ and the canonical extension 
$\widetilde{\beta^n}$ of $\beta^n$ to $\widehat{E_n}$ respectively. Since 
$\widehat{\beta^n}$ commutes with $\widetilde{\beta^n}$, then $d\widehat{\beta^n}+d\widetilde{\beta^n}$ is the infinitesimal generator of $\overline{\beta^n}$. Then we have the following lemma:\\

Lemma 4.5 
\[ (\widehat{P} \otimes I_n)(d\overline{\beta^n})+\widehat{\Omega} \in 
\mathrm{MC}^{(\widehat{M_n(A)},\mathbb{R},\overline{\beta},\widehat{\tau^n})}
(\widehat{M_n(\Xi)}) ,  \]

Proof. \quad Since $\widehat{P} \otimes I_n)(d\overline{\beta^n})$ is the 
Grassmann connection of $\widehat{M_n(\Xi)}$ with respect to the action 
$\overline{\beta}$, Theorem 2.1 implies that it belongs to 
\[{\mathrm{MC}}^{(\widehat{M_n(A)},\mathbb{R},\overline{\beta},\widehat{\tau^n})}(\widehat{M_n(\Xi)})~.\]
\noindent
As $\widehat{\Omega} \in \widehat{E_n}$ is skew adjoint by Lemma 4.4, the 
conclusion follows from Theorem 2.1. ~Q.E.D.  \\

\noindent 
By the Lemma 4.5, we then define a mapping 
\[\Phi_{\beta}:\mathcal{M}^{(M_n(A),\mathbb{R},\beta,\tau^n)}(M_n(\Xi)) 
\longmapsto \mathcal{M}^{(\widehat{M_n(A)},\mathbb{R},\overline{\beta},
\widehat{\tau^n})}(\widehat{M_n(\Xi)})  \]
\noindent
by the following fashionF
\[ \Phi_{\beta}([\nabla]_{U(E_n)}) 
= [(\widehat{P} \otimes I_n)(d\overline{{\beta}^n})+\widehat{\Omega}~]
_{U(\widehat{E_n})},\]
\noindent
where $[\nabla]_{\{*\}}$ means the equivalence class of $\nabla$ under the gauge action of $\{*\}$. \\

We then check the following lemma: \\

Lemma 4.6 \qquad  $\Phi_{\beta}$ is well defined. \\

Proof. \quad Let $\Omega,~\Omega_{1}$ be two skew adjoint elements in $E_n$, 
and suppose $u(\nabla^0_{\beta}+\Omega)u^*=\nabla^0_{\beta}+\Omega_{1}$ 
for some unitary $u \in E_n$, then 
\[\Omega_{1}=u\nabla^0_{\beta}u^*-\nabla^0_{\beta}+u{\Omega}u^* \]
\noindent
We have to show that
$\nabla^0_{\overline{\beta}}+\widehat{\Omega}$ 
is equal to 
$\nabla^0_{\overline{\beta}}+\widehat{\Omega}_{1}$ 
up to the gauge automorphisms of $U(\widehat{E_n})$. Now we compute that\\

$(\nabla_{\widehat{\beta}}^0+\widehat{\Omega}_{1})(\xi)(t)$
\[~~=(\widehat{P} \otimes I_n)(\delta)(\xi)(t)+
(u{\nabla_{\beta}^0}u^*-\nabla_{\beta}^0+u{\Omega}u^*)\{\xi(t)\} \]
\[=2{\pi}it\xi(t)+\widetilde{u}\widehat{\Omega}\widetilde{u}^*(\xi)(t)+
(u{\nabla_{\beta}^0}u^*-\nabla_{\beta}^0)\{\xi(t)\} \]
\noindent
$(\xi \in \widehat{M_n(\Xi)})$ where $\widetilde{u}(\xi)(t)=u\{\xi(t)\}$. 
Since we see that\\

$(u{\nabla_{\beta}^0}u^*-\nabla_{\beta}^0)\{\xi(t)\}$
\[=(\widetilde{u}{\nabla_{\widetilde{\beta}}^0}\widetilde{u}^*-{\nabla_{\widetilde{\beta}}^0})(\xi)(t) ,\]
\noindent
and $\widetilde{u}\nabla_{\widehat{\beta}}^0\widetilde{u}^*(\xi)(t)=2{\pi}it\xi(t)$, then we obtain that 
\[\nabla_{\widehat{\beta}}^0+\nabla_{\widetilde{\beta}}^0+\widehat{\Omega}_{1})(\xi)(t)=\gamma_{\widetilde{u}}(\nabla_{\widehat{\beta}}^0+\nabla_{\widetilde{\beta}}^0+\widehat{\Omega})(\xi)(t) \]
\noindent
$(\xi \in \widehat{M_n(\Xi)})$. As we know that
\[\nabla_{\widehat{\beta}}^0+\nabla_{\widetilde{\beta}}^0
=\nabla_{\overline{\beta}}^0 \],
\noindent
then the conclusion follows.~Q.E.D. \\

By definition, $\widetilde{\beta}$ is a weakly inner action of $\widehat{M_n(A)}$ implimented by a unitary multiplier flow $\mu$ of $\widehat{M_n(A)}$ 
faithfully acting on $L^2(\mathbb{R},H_{\tau^n})$ for the Hilbert space 
$L^2(M_n(A),\tau^n)$. Actually, as $\widetilde{\beta}_t(a)(s)=\beta_t(a(s))$ 
for all $a \in \widehat{M_n(A)},~s,t  \in \mathbb{R}$, then $\mu_t(a)(s)=a(s-t)$ for all $a \in \mathcal{S}(\mathbb{R},M_n(A)),~s,t \in \mathbb{R}$ Then we have the following lemma: \\

Lemma 4.7 \quad The F$^*$-system $(\widehat{M_n(A)},\mathbb{R},\overline{\beta})$ is inner conjugate to the $F^*$-system $(\widehat{M_n(A)},\mathbb{R},\widehat{\beta})$. Then it implies that the $F^*$-system \\

\noindent
$(\widehat{M_n(A)}\rtimes_{\overline{\beta}} \mathbb{R},\mathbb{R},\widehat{\overline{\beta}})$ is isomorphic to the system $(\Doublehat{M_n(A)},\mathbb{R},\Doublehat{\beta})$ \\

\noindent
via the map:
\[\Lambda(x)(t)=\mu_{-t}x(t) \]
\noindent
for all $x \in \widehat{M_n(A)}\rtimes_{\overline{\beta}} \mathbb{R}$, \\

\noindent
where 
\[\Doublehat{M_n(A)}=\widehat{M_n(A)}\rtimes_{\widehat{\beta}} \mathbb{R}~.\]
\noindent
By Lemmas 4.5 and 4.7, we deduce the following lemma: \\

Lemma 4.8  \quad Let $\Lambda_{\beta} : \mathcal{M}^{(\widehat{M_n(A)}\rtimes_{\overline{\beta}} \mathbb{R},\mathbb{R},\overline{\overline{\beta}},\Doublehat{\tau^n})}(\Doublehat{M_n(\Xi)}) \longmapsto $ \\

$\mathcal{M}^{(\Doublehat{M_n(A)},\mathbb{R},\Doublehat{\beta} \circ Ad(\nu),\Doublehat{\tau^n})}(\Doublehat{M_n(\Xi)})$ ~~defined by 
\begin{center}
$\Lambda_{\beta}([\nabla]_{U(\widehat{M_n(A)}\rtimes_{\overline{\beta}} \mathbb{R})})=[(\Lambda^n \circ \nabla \circ (\Lambda^n)^{-1}]_{U(\Doublehat{M_n(A)})}$ \end{center}
\noindent
where $\nu$ is the unitary multiplier flow of $\Doublehat{M_n(A)}$ implimenting  $\widetilde{\widehat{\beta}}$ on $\Doublehat{M_n(A)}$, and $\Lambda^n$ is the 
isomorphism from ${\mathrm{End}}_{\widehat{M_n(A)}\rtimes_{\overline{\beta}} \mathbb{R}}(\Doublehat{M_n(\Xi)})$ onto ${\mathrm{End}}_{\Doublehat{M_n(A)}}(\Doublehat{M_n(\Xi)})$ induced by $\Lambda$. Then it implies that $\Lambda_{\beta}$ 
is a homeomorphism  \\

Proof. \quad By the definition of $\Lambda$, we check that
\[ \Lambda \circ \widehat{\overline{\beta}} \circ 
\Lambda^{-1}=\Doublehat{\beta}~,~\Lambda \circ \widetilde{\overline{\beta}} 
\circ \Lambda^{-1}=\widetilde{\widehat{\beta}}.\]
\noindent
By the same reason as for $\widetilde{\beta}$, there exists a unitary multiplier flow $\nu$ of $\Doublehat{M_n(A)}$ such that $\widetilde{\widehat{\beta}}
=Ad(\nu)$ on $\Doublehat{M_n(A)}$. The rest is easily seen. \quad Q.E.D. \\

Let $\Psi_{\beta}^0$ be the isomorphism from the F$^*$-system $(\Doublehat{M_n(A)},\mathbb{R},\Doublehat{\beta})$ onto the F$^*$-system $(M_n(A) \otimes 
\mathcal{K}^{\infty}(\mathbb{R}),\mathbb{R},\beta \otimes Ad(\lambda))$ defined by
\[ \Psi_{\beta}^0(x)(t,s) = 
			\int_{\mathbb{R}} e^{2{\pi}irs}\beta_{-t}(x(t-s,r))~dr~,  \]
\noindent
and 
\[ (\Psi_{\beta}^0)^{-1}(x)(t,s) = 
		\int_{\mathbb{R}} e^{2{\pi}i(t-r)s}\beta_r(x(r,r-t))~dr   \]
\noindent
$(x \in M_n(A) \otimes \mathcal{K}^\infty(\mathbb{R}),~t,s \in \mathbb{R})$.
By definition, we compute that  \\
$\Psi_{\beta}^0 \circ Ad(\nu_p) \circ (\Psi_{\beta}^0)^{-1}(x)(t,s)$ 
\[~={\int} e^{2{\pi}i(rs+p(t-s))}\beta_{-1}((\Psi_{\beta}^0)^{-1}(x)(t-s,r))~dr~,\]
\noindent
which is equal to 
\[\int\!\!\!\int e^{2{\pi}i(p(t-s)+(t-r')r)}\beta_{r'-t}(x(r',r'-t+s))~dr'dr~.\]\noindent
$(x \in M_n(A) \otimes \mathcal{K}^\infty(\mathbb{R}),~t,s \in \mathbb{R})$. 
Therefore it follows that
\[\Psi_{\beta}^0 \circ Ad(\nu_p) \circ (\Psi_{\beta}^0)^{-1}(x)(t,s)=
   e^{2{\pi}ip(t-s)}x(t,s) ,\]
\noindent
$(x \in M_n(A) \otimes \mathcal{K}^\infty(\mathbb{R}),~p,t,s \in \mathbb{R})$,
 which implies that there exists a unitary multiplier flow $\nu_{\beta}$ of 
$\mathcal{K}^\infty(\mathbb{R})$ with the property that
\[\Psi_{\beta}^0 \circ Ad(\nu_p) \circ (\Psi_{\beta}^0)^{-1}
      =Ad(I \otimes (\nu_{\beta})_p)  \quad (p \in \mathbb{R}) \]
\noindent
on $M_n(A) \otimes \mathcal{K}^\infty(\mathbb{R})$. Then it turns out that
\[ \Psi_{\beta}^0 \circ (\Doublehat{\beta} \circ Ad(\nu)) \circ (\Psi_{\beta}^0)^{-1}=\beta \otimes Ad(\lambda \circ \nu_{\beta}) ~.\]
\noindent
Let $\Psi_{\beta}$ be the map:
$\mathcal{M}^{(\Doublehat{M_n(A)},\mathbb{R},\Doublehat{\beta} \circ Ad(\nu),\Doublehat{\tau^n})}(\Doublehat{M_n(\Xi)}) \longmapsto $ \\

$\mathcal{M}^{(M_n(A) \otimes {\mathcal{K}}^\infty(\mathbb{R}),\mathbb{R},\beta \otimes Ad(\lambda \circ \nu_{\beta}),{\tau^n} \otimes Tr)}(M_n(\Xi) \otimes {\cal{K}}^\infty(\mathbb{R})) $ \\

\noindent
induced by the equivariant isomorphism $\Psi^0_{\beta}$. Then we also show the 
following lemma by the same way as Lemma 4.8: \\

Lemma 4.9 \quad  $\Psi_{\beta}$ is a homeomorphism induced by the equivariant 
isomorphism $\Psi_{\beta}^0$. \\

Let us now consider the following map: 
\[\Pi_{\beta} : \mathcal{M}^{(M_n(A) \otimes {\mathcal{K}}^\infty(\mathbb{R}),\mathbb{R},\beta \otimes Ad(\lambda \circ \nu_{\beta}),{\tau^n} \otimes Tr)}(M_n(\Xi) \otimes \mathcal{K}^\infty(\mathbb{R})) \]
$~~~~~~~~~\longmapsto \mathcal{M}^{(M_n(A),\mathbb{R},\beta,{\tau}^n)}(M_n(\Xi))$ \\ 

defined by the natural one induced from the map $\Pi:$
\[M_n(A) \otimes \mathcal{K}^\infty(\mathbb{R}) \longmapsto M_n(A) \]
\noindent
given by $\Pi:x \mapsto (I \otimes e)x(I \otimes e)$, 
where $e$ is a rank one projection of $\mathcal{K}^\infty(\mathbb{R})$,~
$Tr$ the canonical trace of $\mathcal{K}^\infty(\mathbb{R})$.
\noindent
Now let $\nabla \in \rm{MC}(M_n(\Xi) \otimes \mathcal{K}^\infty(\mathbb{R}))$ 
and put 
\[  \nabla_e(\xi)=(I_n \otimes e)\nabla(\xi \otimes e)  \]
\noindent
$(\xi \in M_n(\Xi))$, 
where $I_n$ is the identity of $E_n$. Then $\nabla_e$ is well defined and 
independent of the choice of $e$ up to the gauge equivalence 
because of the existence of a unitary multiplier of $\mathcal{K}^\infty(\mathbb{R}))$ which sends $e$ to another rank one projection. Then we see that 
given any $u \in U(E_n \otimes \mathcal{K}^\infty(\mathbb{R}))$,
\[  (\gamma_u(\nabla))_e =\gamma_{u_e}(\nabla_e)  , \]
where $u_e = (I_n \otimes e)u(I_n \otimes e) \in U(E_n)$. Here we define a map 
$\Pi_{\beta}$ by 
\[\Pi_{\beta}([\nabla]_{U(E_n \otimes \mathcal{K}^\infty(\mathbb{R}))}
=[\nabla_e]_{U(E_n)}=[\Pi^n \circ \nabla \circ (\Pi^n)^{-1}]_{U(E_n)} ~.\]
Then it is well defined and independent of the choice of $e$. Moreover, we have the following lemma:  \\

Lemma 4.10 
\[\Pi_{\beta}:\mathcal{M}^{(M_n(A) \otimes \mathcal{K}^\infty(\mathbb{R}),\mathbb{R},\beta \otimes Ad(\lambda \circ \nu_{\beta})),{\tau^n} \otimes Tr)}(M_n(\Xi) \otimes \mathcal{K}^\infty(\mathbb{R})) \]
$~~~~~~~\longmapsto \mathcal{M}^{(M_n(A),\mathbb{R},\beta,{\tau}^n)}(M_n(\Xi))$~is a homeomorphism. \\ 

Proof. \quad Let us define the mapping $\Pi^{-1}_{\beta}$ by
\[\Pi^{-1}_{\beta}([\nabla]) ~=~ [\nabla \otimes I_{\mathcal{K}^\infty(\mathbb{R})}]   ~.   \]
Then it is easily seen that both $\Pi^{-1}_{\beta} \circ \Pi_{\beta}$ and 
$\Pi_{\beta} \circ \Pi^{-1}_{\beta}$ are identities. Moreover if 
$[\nabla^{\iota}] \longrightarrow [\nabla]$ with respect to 
$< \cdot \mid \cdot >_{M_n(A)}$, then it follows from the definition that 
there exists a unitary net $\{u_{\iota}\}$ (by choosing a subnet) of $E_n$ 
such that $\gamma_{u_{\iota}}(\nabla^{\iota}) \longrightarrow \nabla$, 
which implies that $[(\nabla^{\iota})_e] \longrightarrow [\nabla_e]$, 
so that $\Pi_{\beta}$ is continuous and so is also $\Pi^{-1}_{\beta}$ by the 
same way. ~Q.E.D.  \\

Let $\nabla^0_{\cdot}$ be the Grassmann connection of $\cdot$. Then we easily 
check the following lemma: \\

Lemma 4.11
\[ \quad \Psi^0_{\beta^n} \circ \nabla^0_{\Doublehat{\beta} \circ Ad(\nu)} \circ (\Psi^0_{\beta^n})^{-1} =\nabla^0_{\beta \otimes Ad(\lambda \circ \nu_{\beta})}   ~.  \]  
Proof. \quad It follows from Lemma 4.9 that
\[  \Psi^0_{\beta} \circ (\Doublehat{\beta} \circ Ad(\nu) \circ (\Psi^0_{\beta})^{-1} = \beta \otimes Ad(\lambda \circ \nu_{\beta}) ~ . \]
\noindent
Since
\[ \Doublehat{M_n(\Xi)} = \mathcal{S}(\mathbb{R}^2,M_n(\Xi)) , \] 
and 
\[ \nabla^0_{\Doublehat{\beta}} =\Doublehat{P}~d~\Doublehat{\beta^n} ,\]
where 
\[ \Doublehat{P}=
   P \otimes I_n \otimes I_{\mathcal{S}(\mathbb{R}^2)} ,  \]
\noindent
and $d~\Doublehat{\beta^n}$ is the infinitesimal generator of 
$\Doublehat{\beta^n}$, then this implies the conclusion. ~Q.E.D. \\

\noindent
Moreover, we need the following lemma which is directly shown: \\

Lemma 4.12  \quad There exists a 
\[U \in U(\mathrm{End}_{M_n(A) \otimes \mathcal{K}^{\infty}(\mathbb{R})}(M_n(\Xi) \otimes {\mathcal{S}}(\mathbb{R}^2))) \]
such that
\[ \Psi^0_{\beta^n} \circ \Lambda^n \circ \Doublehat{\Omega} \circ (\Psi^0_{\beta^n}\circ \Lambda^n)^{-1}=\gamma_{U}(\Omega \otimes Id) \]  
\noindent
Actually, $U$ is defined as $U(\xi)(s,t)=u^n_{-s}\xi(s,t)$ 
for all $\xi \in M_n(\Xi) \otimes {\mathcal{S}}(\mathbb{R}^2)$, 
where $u^n$ is the unitary flow to $E_n$ induced by $\beta^n$. \\

Proof. \quad We know by definition that
\[ \Psi^0_{\beta^n}(\xi)(s,t) ~= 
	 {\int}_{\mathbb{R}}~e^{2{\pi}irt}~u^n_{-s}\xi(s-t,r)~dr   \]
\noindent
$(\xi \in \Doublehat{M_n(\Xi)}$, and
\[ (\Psi^0_{\beta^n})^{-1}(\xi)(s,t) ~= 
	\int_{\mathbb{R}}~e^{2{\pi}i(s-r)t}~u^n_r\xi(r,r-s)~dr   \]
\noindent
$(\xi \in M_n(\Xi) \otimes \mathcal{S}({\mathbb{R}}^{2}))$. Then we compute that
\[ \Psi^0_{\beta^n} \circ \Lambda^n \circ \Doublehat{\Omega}(\xi)(s,t)=
\int~e^{2{\pi}irt}u^n_{-s}(\Lambda^n \circ \Doublehat{\Omega}(\xi)(s-t,r))~dr \]\noindent
$(\xi \in \Doublehat{M_n(\Xi)})$. Then as we know that
\[\Lambda^n \circ \Doublehat{\Omega}(\xi)(s-t,r)=\nu^n_{-r}\widehat{\Omega}\{\xi(r)\}(s-t)=\Omega\{\xi(s-t+r,r)\} ~.\]
Therefore we have that \\

$ \Psi^0_{\beta^n} \circ \Lambda^n \circ \Doublehat{\Omega}(\xi)(s,t) $
\[ ~=~\int~e^{2{\pi}irt}u^n_{-s}\Omega\{(\xi)(s-t+r,r)\}~dr \]
\noindent
$(\xi \in \Doublehat{M_n(\Xi)})$. Replacing $\xi$ by $(\Psi^0_{\beta^n} \circ 
\Lambda^n)^{-1}(\xi)$, we obtain that  \\

$(\Psi^0_{\beta^n} \circ \Lambda^n)^{-1}(\xi)(s-t+r,r)$ 
\[~~~=\int_{\mathbb{R}}~e^{2{\pi}i(s-t-r')r}u^n_{r'}\xi(r',r'-s+t)~dr' \]
\noindent
$(\xi \in M_n(\Xi) \otimes \mathcal{S}(\mathbb{R}^2)$. Combining the argument 
discussed above, we deduce that \\

$(\Psi^0_{\beta^n} \circ \Lambda^n) \circ {\Doublehat{\Omega}} \circ (\Psi^0_{\beta^n} \circ \Lambda^n)^{-1}(\xi)(s,t)$ 
\[=\int\!\!\!\int~e^{2{\pi}i(s-r')r}u^n_{-s}\Omega\{u^n_{r'}\xi(r',r'-s+t)\}~dr'dr ~,\]
\noindent
which is equal to 
\[u^n_{-s}\Omega\{u^n_s\xi(s,t)\}=\gamma_{U}(\Omega \otimes Id)(\xi)(s,t)~,\]
\noindent
where $U(\xi)(s,t)=u^n_{-s}\xi(s,t)$. This implies the conclusion.~Q.E.D. \\
\noindent
We next show the following lemma which seems to be essential to prove our main 
theorem: \\

Lemma 4.13
\[ \Pi_{\beta} \circ \Psi_{\beta} \circ \Lambda_{\beta} \circ \Phi_{\overline{\beta}} \circ \Phi_{\beta} = Id   \]
\noindent
on $\mathcal{M}^{(M_n(A),\mathbb{R},\beta,\tau^n)}(M_n(\Xi))$, 
where $\Psi_{\beta}$ is the homeomorphism on the moduli space induced from the 
isomorphism $\Psi^0_{\beta}$ given in Lemma 4.9. \\

Proof. \quad Let $\nabla=\nabla^0_{\beta^n}+\Omega \in \mathrm{MC}^{(M_n(A),\mathbb{R},\beta,\tau^n)}(M_n(\Xi))$. 
Then we know by Lemmas 4.11 and 4.12 that there exist a 
\[U \in U(\mathrm{End}_{M_n(A) \otimes {\mathcal{K}}^{\infty}(\mathbb{R})}(M_n(\Xi) \otimes \mathcal{S}(\mathbb{R}^2))) \]
such that
\[(\Psi^0_{\beta^n} \circ \Lambda^n) \circ \nabla^0_{\overline{\widehat{\beta}}} \circ (\Psi^0_{\beta^n} \circ \Lambda^n)^{-1} = \nabla^0_{\beta \otimes Ad(\lambda \circ \nu_{\beta})}  ,  \]
\noindent
and 
\[(\Psi^0_{\beta^n} \circ \Lambda^n) \circ \Doublehat{\Omega} \circ (\Psi^0_{\beta^n} \circ \Lambda^n)^{-1} = \gamma_{U}(\Omega \otimes Id)  ~.  \]  
\noindent
By Lemmas 4.11 and 4.12, we obtain that \\

$\Psi_{\beta} \circ \Lambda_{\beta} \circ \Phi_{\overline{\beta}} \circ \Phi_{\beta}([\nabla]_{U(E_n)})$ 
\[=[(\nabla^0_{\beta^n \otimes Ad(\lambda \circ \nu_{\beta})})+\gamma_{U}(\Omega \otimes Id)]_{U(E_n \otimes \mathcal{K}^{\infty}(\mathbb{R}))}~. \]
\noindent
then we see that
\[(\nabla^0_{\beta^n \otimes Ad(\lambda \circ \nu_{\beta})})_e=\nabla^0_{\beta^n } \otimes e   ~,\]
\noindent
where $e$ is a rank one projection of $\mathcal{K}^{\infty}(\mathbb{R})$. In fact, 
we check that  \\

$ (P \otimes I_n \otimes e)d((\beta^n) \otimes Ad(\lambda \circ \nu_{\beta}))^n(\xi \otimes e) $
\[=(P \otimes I_n \otimes e)\{d((\beta^n)^n \otimes \iota)+
d(\iota \otimes Ad(\lambda \circ \nu_{\beta})^n)\}(\xi \otimes e)~ . \]
\noindent
$(\xi \in M_n(\Xi))$. Then it follows that
\[ e{Ad}(\lambda_t \circ (\nu_{\beta})_t)(e)=e  . \]
\noindent
for all $t \in \mathbb{R}$. Therefore, it follows that
\[(P \otimes I_n \otimes e)d(Id \otimes Ad(\lambda \circ \nu_{\beta})^n)(\xi \otimes e)=0 \]
\noindent
$(\xi \in M_n(\Xi))$. On the other hand, we know that
\[ \gamma_{U}(\Omega \otimes Id)_e = \gamma_{U_e}(\Omega) \otimes e ~,\]
\noindent
where $U_e$ belongs to $U(E_n)$ with $U_e \otimes e=(I \otimes e)U(I \otimes e)$. By the definition of $U$, $U_e$ commutes with $\beta^n$. Consequently, 
it follows that
\[ [\nabla^0_{\beta^n \otimes e}+\gamma_{U_e}(\Omega) \otimes e]_{U(E_n) \otimes e}=[\nabla^0_{\beta^n}+\Omega]_{U(E_n)} , \]
\noindent
which deduce the conclusion. ~Q.E.D.  \\

\noindent
Applying Lemma 4.13 to the system 
$(\widehat{M_n(A)},\mathbb{R},\overline{\beta})$, we obtain the following 
corollary: \\

Corollary 4.14 
\[\Pi_{\overline{\beta}} \circ \Psi_{\overline{\beta}} \circ 
\Lambda_{\overline{\beta}} \circ \Phi_{\overline{\overline{\beta}}} \circ 
\Phi_{\overline{\beta}} = \mathrm{Id} \]
\noindent
holds on 
\[ \mathcal{M}^{(\widehat{M_n(A)},\mathbb{R},\overline{\beta},\widehat{\tau^n})}(\widehat{M_n(\Xi)})~, \]
\noindent
where the map $\Pi_{\overline{\beta}}$ is a homeomorphism from \\

$\mathcal{M}^{(\widehat{M_n(A)} \otimes \mathcal{K}^\infty(\mathbb{R}),\mathbb{R},\overline{\beta} \otimes Ad(\lambda \circ \nu_{\overline{\beta}}),\widehat{\tau^n} \otimes Tr)}(\widehat{M_n(\Xi)} \otimes \mathcal{K}^\infty(\mathbb{R}))$ \\
\noindent
onto \\

${\mathcal{M}}^{(\widehat{M_n(A)},\mathbb{R},\overline{\beta},\widehat{\tau^n})}(\widehat{M_n(\Xi)})$ ~~induced by the mapping:
\[ \widehat{M_n(A)} \otimes \mathcal{K}^\infty(\mathbb{R}) \longmapsto 
\widehat{M_n(A)}  \]
\noindent
given by $x \mapsto (I \otimes e)x(I \otimes e)~. $ \\

As we have seen in Lemma 4.8, 
\[\Lambda \circ \overline{\overline{\beta}} \circ \Lambda^{-1}=
\Doublehat{\beta} \circ Ad(\nu) \]
Then the relation between $\Phi_{\Doublehat{\beta} \circ Ad(\nu)}$ and 
$\Phi_{\beta}$ is as followsF\\

Lemma 4.15
\[ \Phi_{\Doublehat{\beta} \circ Ad(\nu)}=\Lambda_{\beta} \circ (\Pi_{\overline{\beta}} \circ \Psi_{\overline{\beta}} \circ \Lambda_{\overline{\beta}})^{-1} 
\circ \Phi_{\beta} \circ \Pi_{\beta} \circ \Psi_{\beta}   \]
\noindent
holds on 
\[\mathcal{M}^{(\Doublehat{M_n(A)},\mathbb{R},\Doublehat{\beta} \circ Ad(\nu),
\Doublehat{\tau^n})}(\Doublehat{M_n(\Xi)}) ~.\]

Proof. \quad By Lemmas 4.8 $\sim$ 4,13 and Corollary 4.14, we have that 
\[ (\Pi_{\overline{\beta}} \circ \Psi_{\overline{\beta}}\circ \Lambda_{\overline{\beta}})^{-1} = \Phi_{\overline{\overline{\beta}}} \circ \Phi_{\overline{\beta}}~. \]
\noindent
By the equality written just above this Lemma, we see that
\[\Lambda_{\beta} \circ \Phi_{\overline{\overline{\beta}}} \circ (\Lambda_{\beta})^{-1}=\Phi_{\Doublehat{\beta} \circ Ad(\nu)} ~.\]
\noindent
By Lemma 4.13 and Corollary 4.14, $\Phi_{\overline{\beta}}$ is bijective. 
By Lemma 4.12, we have that 
\[ \Phi_{\overline{\beta}}^{-1} = \Phi_{\beta} \circ \Pi_{\beta} \circ \Psi_{\beta} ~,  \]
\noindent
which implies the conclusion.  ~Q.E.D.  \\
\noindent
Summing up the above argument, we obtain the following lemma: \\

Lemma 4.16 \quad $\Phi_{\beta}$ and $\Phi_{\overline{\beta}}$ are 
homeomorphisms. \\

Proof. \quad By Lemmas 4.8 $\sim$ 4.15, $\Phi_{\beta}$ and $\Phi_{\overline{\beta}}$ are bijective and bicontinuous, which completes the proof. \quad Q.E.D.\\

\noindent
We then define a map $(\Phi_{\alpha^n})^{-1}$:
\[\mathcal{M}^{(M_n(A) \rtimes_{\alpha^n} \mathbb{R}, \mathbb{R}, \overline{\alpha^n}, \widehat{\tau^n})}(\widehat{M_n(\Xi)}) \]
\[\longmapsto \mathcal{M}^{(M_n(A),\mathbb{R},\alpha^n,\tau^n)}(M_n(\Xi)) \]
\noindent
defined by 
\[\Pi_{\alpha^n} \circ \Psi_{\alpha^n} \circ (\widetilde{\iota})^{-1} \circ \Lambda_{\beta} \circ \Phi_{\overline{\beta}} \circ \iota ~,\]
\noindent
where $\iota$ is the extended map on $\mathcal{M}^{(M_n(A),\mathbb{R},\alpha^n,\tau^n)}(M_n(\Xi))$ induced by $\iota_u$ in Lemma 4.3, and so is $\widetilde{\iota}$ on $\mathcal{M}^{(\widehat{M_n(A)},\mathbb{R},\overline{\alpha^n},\widehat{tau^n})}(\widehat{M_n(\Xi)})$ induced by $\widetilde{\iota_u}$, 
because the map $\iota_u$ intertwines $\overline{\alpha^n}$ and $\overline{\beta}$, and the map $\widetilde{\iota_u}$ intertwines $\overline{\overline{\alpha^n}}$ and $\overline{\overline{\beta}}$.   \\

Lemma 4.17 \quad $(\Phi_{\alpha^n})^{-1}$ is a homeomorphism. \\

Proof. \quad By Lemma 4.3 (2), there exists a unitary multiplier $W$ of 
$M_n(A) \otimes \mathcal{K}^\infty(\mathbb{R})$ such that 
\[Ad(W) \circ \Psi_{\beta}^0 \circ \widetilde{\iota_u} = \Psi_{\alpha^n}^0~,\]
\noindent
which implies that $\Psi_{\beta} \circ \widetilde{\iota} = \Psi_{\alpha^n}$. 
Moreover, $\iota$ and $\widetilde{\iota}$ are homeomorphisms. Then it follows 
from Lemma 4.16 that $(\Phi_{\alpha^n})^{-1}$ is a homeomorphism. ~Q.E.D.\\
\noindent
By Lemmas 4.17, we deduce the following corollary: \\

Corollary 4.18 \quad $\Phi_{\alpha^n}$ is a homeomorphism:
\[\mathcal{M}^{(M_n(A),\mathbb{R},\alpha^n,\tau^n)}(M_n(\Xi)) \longmapsto 
\mathcal{M}^{(M_n(A) \rtimes_{\alpha^n} \mathbb{R}, \mathbb{R}, \overline{\alpha^n}, \widehat{\tau^n})}(\widehat{M_n(\Xi)})~.\]

\noindent
Finally, using $\Phi_{\alpha^n}$, we define the map $\Phi_{\alpha}$ : 
\[\mathcal{M}^{(A,\mathbb{R},\alpha,\tau)}(\Xi) \longmapsto \mathcal{M}^{(\widehat{A},\mathbb{R},\overline{\alpha},\widehat{\tau})}(\widehat{\Xi}) \]
\noindent
by
\[\Phi_{\alpha}=\widehat{\Pi_n} \circ \Phi_{\alpha^n} \circ (\Pi_n)^{-1} ~,\]
\noindent 
where $\Pi_n$ is a homeomorphism:
\[\mathcal{M}^{(A \otimes M_n(\mathbb{C}),\mathbb{R},\alpha \otimes Ad(\lambda \circ \nu_{\alpha}),\tau \otimes Tr_n)}(\Xi \otimes M_n(\mathbb{C})) \longmapsto \mathcal{M}^{(A,\mathbb{R},\alpha,\tau)}(\Xi) ~.\]
\noindent
Finally, we show the following main lemma: \\

Lemma 4.19  \quad $\Phi_{\alpha}$ is a homeomorphism. \\

Proof.  In Lemma 4.10, replacing $(M_n(A) \otimes \mathcal{K}^\infty(\mathbb{R}),\beta \otimes Ad(\lambda \circ \nu_{\beta})$ and $M_n(\Xi) \otimes \mathcal{K}^{\infty}(\mathbb{R})$ by $A \otimes M_n(\mathbb(C)),\alpha^n$ and $\Xi \otimes 
M_n(\mathbb{C})$ respectively, we deduce that both $\widehat{\Pi_n}$ and 
$(\Pi_n)^{-1}$ are homeomorphisms. Then it implies the conclusion 
by Corollary 4.18. ~Q.E.D.  \\

\noindent
Summing up all the argument discussed above, we obtain the main result of 
Theorem 4.1  \\

In what follows, we compute the moduli spaces of some concrete examples 
by means of Theorem 4.1 : \\

Example 4.20  
\[\mathcal{M}^{(\mathcal{K}^{\infty}(\mathbb{R}),\mathbb{R},Ad(\lambda),Tr)}
({\mathcal{K}}^{\infty}(\mathbb{R})) \approx {\mathcal{M}}^{({\mathcal{S}}(\mathbb{R}),\mathbb{R},\lambda,\int)}(\mathcal{S}(\mathbb{R}))  \]
\[ \qquad~~~~~~~~~~~~~\qquad ~~~~~\quad ~~\approx {\mathcal{M}}^{(\mathbb{C},\mathbb{R},Id,1)}(\mathbb{C}) \approx \mathbb{R} , \]
where $\approx$ means a symbol of homeomorphism.  \\

Examples 4.21 \quad Given a $\theta \in \mathbb{R}$, let us take the Moyal 
product $\star_{\theta}$ on $\mathcal{S}(\mathbb{R}^2)$. Then $(\mathcal{S}(\mathbb{R}^2),\star_{\theta})$ becomes a $F^*$-algebra, which is denoted by $\mathbb{R}_{\theta}^2$. Since $\mathbb{R}_{\theta}^2$ is isomorphic to $\mathcal{S}(\mathbb{R}) \rtimes_{\theta} \mathbb{R}$, then it follows from Theorem 4.1 that
\[{\cal{M}}^{(\mathbb{R}_{\theta}^2,\mathbb{R},\overline{\theta},\tau_{\theta})}(\mathbb{R}_{\theta}^2) \approx \mathbb{R}  , \]
\noindent
where $\tau_{\theta}$ is the canonical trace of $\mathbb{R}_{\theta}^2$. \\

\noindent
Even though changing F$^*$-flows into F$^*$-multiflows, the same result as 
Theorem 4.1 is obtained by applying it repeatedly. Actually, we now take a $F^*$-dynamical system $(A,\mathbb{R}^2,\alpha)$ with a faithful $\alpha$-invariant 
trace $\tau$ on $A$. Let $\alpha^i_t=\alpha_{(t,0)}$ if $i=1$, $=\alpha_{(0,t)}$ if $i=2$. By definition, $\alpha^1$ 
commutes with $\alpha^2$ and $\alpha=\alpha^1 \circ \alpha^2$. Let us take 
$\widehat{A}^1=A \rtimes_{\alpha^1} \mathbb{R}$. Then we know that $\widehat{A}=\widehat{A}^1 \rtimes_{\tilde{\alpha^2}} \mathbb{R}$. We now apply Theorem 4.1 
to the $F^*$-dynamical  system $(A,\alpha^1,\mathbb{R})$, $\Xi$ and $\tau$. Then it follows that
\[{\mathcal{M}}^{(A,\mathbb{R},\alpha^1,\tau)}(\Xi) \approx {\mathcal{M}}^{(\widehat{A}^1,\mathbb{R},\overline{\alpha^1},\widehat{\tau}^1)}(\widehat{\Xi}^1)~.\]\noindent
where $\widehat{\tau}^1$ is the dual trace of $\tau$ on $\widehat{A}^1$ and $\widehat{\Xi}^1 \otimes_{A} \widehat{A}^1.$ Since $\alpha^1$ commutes with 
$\alpha^2$, it follows from definition that $\overline{\alpha^1}$ commutes with $\tilde{\alpha^2}$. Then by the same way as the proof of Theorem 4.1 again, we 
obtain that 
\[{\mathcal{M}}^{(A,\mathbb{R}^2,\alpha,\tau)}(\Xi) \approx {\mathcal{M}}^{(\widehat{A}^1,\mathbb{R}^2,\overline{\alpha^1} \circ \tilde{\alpha^2},\widehat{\tau}^1)}(\widehat{\Xi}^1)~.\]
\noindent
In fact, the method used in the main proof of Theorem 4.1 takes place only on 
the $F^*$-dynamical systems $(A,\mathbb{R}^2,\alpha)$ and $(\widehat{A}^1,\mathbb{R}^2, \overline{\alpha^1},\circ \tilde{\alpha^2})$, so that there is no affection from the action $\alpha^2$. Actually, the moduli map from the former one to the latter one is defined by 
\[ [d\alpha^n + \Omega_{\alpha}] ~\longrightarrow ~[d(\overline{\alpha^1} \circ \widetilde{\alpha^2})^n + \widetilde{\Omega_{\alpha}}]~.\] 
\noindent
Then the next step is to be shown that 
\[{\mathcal{M}}^{(\widehat{A}^1,\mathbb{R}^2,\overline{\alpha^1} \circ 
\widetilde{\alpha^2},\widehat{{\tau}^1})}(\widehat{\Xi}^1) \approx 
{\mathcal{M}}^{(\widehat{A}^1 \rtimes_{\tilde{\alpha^2}} \mathbb{R},\mathbb{R}^2,\overline{\tilde{\alpha^2}} \circ \widetilde{\overline{\alpha^1}},\widehat{\tau})}(\widehat{\Xi})~.\]
\noindent
where $\widehat{\Xi}=\Xi \otimes_A A \rtimes_{\alpha} \mathbb{R}^2$ and 
$\widehat{\tau}$ is the dual trace of $\tau$ on $A \rtimes_{\alpha} \mathbb{R}^2$. Under the isomorphism $\Phi$  from $\widehat{A}^1 \rtimes_{\tilde{\alpha^2}} \mathbb{R}$ onto $A \rtimes_{\alpha} \mathbb{R}^2$, we then check that 
the action 
$\overline{\widetilde{\alpha^2}} \circ \widetilde{\overline{\alpha^1}}$ is 
equivalent to $\overline{\alpha}~.$ Actually the former action is transformed 
into $\overline{\alpha^2} \circ \overline{\alpha^1}$ of $\mathbb{R}^2$ on 
$A \rtimes_{\alpha} \mathbb{R}^2$ by taking $Ad(\Phi)$, which is nothing but 
the latter action. We then obtain the following theorem: \\

Theorem 4.22  \quad Let $(A,\mathbb{R}^n,\alpha)~(n \geq 2)$ be a 
F$^*$-multiflow with a faithful continuous $\alpha$-invariant trace $\tau$, and $\Xi$ a finitely generated projective $A$-module. Then there exist a 
F$^*$-multiflow $(\widehat{A},\mathbb{R}^n,\overline{\alpha})$ with a dual trace $\widehat{\tau}$, and a dual $\widehat{A}$-module $\widehat{\Xi}$ such that 
\[{\mathcal{M}}^{(A,\mathbb{R}^n,\alpha,\tau)}(\Xi) \approx 
{\mathcal{M}}^{(\widehat{A},\mathbb{R}^n,\overline{\alpha},\widehat{\tau})}(
\widehat{\Xi})~. \]
\noindent
where $\widehat{A}=A \rtimes_{\alpha} \mathbb{R}^n,~\overline{\alpha}=\widehat{\alpha} \circ \widetilde{\alpha},~\widehat{\Xi}=\Xi \otimes_A \widehat{A}$, and 
$\widehat{\tau}$ is the dual trace of $\tau$ on $\widehat{A}$.\\

Remark. \quad In multiflow cases, the curvatures of Yang-Mills connections 
are non-zero 2-tensors in general although they always vanish in single flow 
cases. \\

\noindent
Using the Theorem 4.22, the similar statement as Corollary 4.2 is obtained in 
the following: \\

Corollary 4.23 \quad Let $(A,\mathbb{R}^n,\alpha)$ be a F$^*$-multiflow with a 
faithful continuous $\alpha$-invariant trace $\tau$, and $(\widehat{A},\mathbb{R}^n,\overline{\alpha})$ its associated F$^*$-flow with the dual trace $\widehat{\tau}$. Suppose $(\widehat{A},\mathbb{R}^n,\beta)$ is another $F^*$-multiflow such that 
\[\widehat{\tau} \cdot \beta = \widehat{\tau} , ~\beta \circ \overline{\alpha} = \overline{\alpha} \circ \beta ~\]
\noindent
then given a finitely generated projective $\widehat{A}$-module $\Xi$, there exist a F$^*$-multiflow $(A,\mathbb{R}^n,\beta_{A})$, a finitely generated projective $A$-module $\Xi_{A}$ and a faithful $\beta_{A}$-invariant trace $\tau_{A}$ of $A$ such that 
\[\mathcal{M}^{(\widehat{A},\mathbb{R}^n,\beta,\widehat{\tau})}(\Xi) \approx 
 \mathcal{M}^{(A,\mathbb{R}^n,\beta_{A},\tau_{A})}(\Xi_{A})  .  \]

\Large{\S5.~Application} \large \quad In this section, we apply Theorem 4.22 and Corollary 4.23 to compute the moduli space in the case of the instanton bundles on the noncommutative Euclidean 4-space with respect to the canonical space 
translations without using the ADHM construction. Actually, it appears as a 
Higgs branch of the theory of D0-branes bound to D4-branes by the expectation 
value of the B-field as well as a regularized version of the target space of 
supersymmetric quantum mechanics arising in the light cone description of (2,0) superconformal theories in six dimensions, although its algebraic structure has already been established in the example 10.1 of [12] (cf.[7],[8]): \\

Let $\mathbb{R}^4_{\theta}$ be the noncommutative $\mathbb{R}^4$ for an 
antisymmetric 4x4 matrix $\theta = (\theta_{i,j})$, in other words, 
$\mathbb{R}^4_{\theta}$ is the $F^*$-algebra generated by 4-selfadjoint elements $\{x_i\}_{i=1}^4$ with the property that \\

\noindent
$(1)~~~~~~~~~~~~~~~~~ [x_i,x_j] = \theta_{i,j}     $ \\

\noindent
$(i,j=1,\cdots,4)$. In other words, 
\[ \mathbb{R}^4_{\theta}=\{ \int_{\mathbb{R}^4} f(t_1,t_2,t_3,t_4)~x_1^{2{\pi}it_1}x_2^{2{\pi}it_2}x_3^{2{\pi}it_3}x_4^{2{\pi}it_4}~dt_1dt_2dt_3dt_4~|~f \in S(\mathbb{R}^4)~\}  \]
\noindent
as a $F^*$-algebra, where $S(\mathbb{R}^4)$ is the set of all rapidly decreasing complex valued functions on $\mathbb{R}^4$. Let $x^i=\theta^{i,j}x_j~(i,j=1,\cdots,4)$ where $(\theta^{i,j})$ is the inverse matrix of $(\theta_{i,j})$. Then 
the $F^*$-algebra $\mathbb{R}^4_{\theta}$ depends essentially on one positive 
real number denoted by the same symbol $\theta$, which satisfy the following 
relation: \\

\noindent
$(2) ~~~~~~~~~ [z_i^*,z_i] = \theta~,~[z_i,z_j]=[z_i^*,z_j]=0 ~~(i,j=0,1,i\neq j) $ \\

\noindent
where $z_0=x^1+\sqrt{-1}x^2,z_1=x^3+\sqrt{-1}x^4$ and $z_i^*$ are the conjugate operators of $z_i$. Let us consider the canonical action $\alpha$ of 
$\mathbb{R}^4$ on $\mathbb{R}^4_{\theta}$ defined by \\

\noindent
$(3)~~~~~~~~~~~~~~~~~ \alpha_{t_i}(x_i) = x_i + t_i     $ \\

\noindent
$(t_i \in \mathbb{R},i=1,\cdots,4)$. Then it is easily seen that the triplet 
$(\mathbb{R}^4_{\theta},\mathbb{R}^4,\alpha)$ is a $F^*$-dynamical system, and 
we easily see that \\

\noindent
$(4)~~~~~~~~~~~~~~~~~ \alpha_{w_i}(z_i)= z_i + w_i     $ \\

\noindent
$(w_i \in \mathbb{C}, i=0,1)$. 
By $(2)$, $\mathbb{R}^4_{\theta}$ is nothing but the $F^*$-tensor product $A_0 
\otimes A_1$ where $A_i$ are the $F^*$-algebras generated by $z_i~(i=0,1)$. 
We now check the algebraic structure of $A_i$. By $(2)$, it follows from [8](cf.[11]) that there exist two Fock spaces $H_i$ such that \\

$~~~~~~z_i(\xi^i_n)=\sqrt{(n+1)\theta}~\xi^0_{n+1}~,~z_i^*(\xi^i_n)=\sqrt{n \theta}~\xi^i_{n-1}$, \\

\noindent
where $\{\xi^i_n\}$ are complete orthonormal systems of $H_i$ 
with respect to the following inner product:
\[ <~f~|~g~>=\sum (n+1)\theta~f(n)\overline{g(n)} \]
\noindent
for two $\mathbb{C}$-valued functions $f,g$ on $\mathbb{N}$ such that  
\[ \sum (n+1)\theta~|f(n)|^2 < \infty~,~\sum (n+1)\theta~|g(n)|^2 < \infty ~.\]
\noindent
for $i=0,1$. 
We may assume that the $A_i$ act on $H_i$ irreducibly. Then it also follows 
from [11] that the $F^*$-algebras $A_i$ are isomorphic to the $F^*$-algebras 
$\mathcal{K}^{\infty}(H_i)$ defined by
\[  \mathcal{K}^{\infty}(H_i)=\{T \in \mathcal{K}(H_i)~|~\{\lambda_k\} \in 
S(\mathbb{N}) \}  \]
\noindent
where $\{\lambda_k\}$ are all eigen values of $T$ and $S(\mathbb{N})$ are the 
set of all sequences $\{c_n\}$ of $\mathbb{C}$ with 
$sup_{n\geq 1}~(1+|n|)^k|c_n| < \infty $ for all $k \geq 0$. Therefore, 
the $F^*$-algebra $\mathbb{R}^4_{\theta}$ is isomorphic to $\mathcal{K}^{\infty}(H_0 \otimes H_1)$. We then have the following proposition: \\

Proposition 1~(cf:[12]). \quad If $\theta \neq 0$, then $\mathbb{R}^4_{\theta}$ is isomorphic to $\mathcal{K}^{\infty}(L^2(\mathbb{C}^2))$ as a $F^*$-algebra.\\
\noindent
By the above Proposition, $\mathcal{K}^{\infty}(L^2(\mathbb{C}^2))$ is the $F^*$-crossed product $S(\mathbb{C}^2) \rtimes_{\tau} \mathbb{C}^2$ of $S(\mathbb{C}^2)$ by the shift action $\tau$ of $\mathbb{C}^2$. We then consider the action 
$\alpha$ defined before. By $(4)$, it follows from [R] that $\alpha$ plays a 
role of the dual action of $\tau$. Then the $F^*$-crossed product $\widehat{\mathbb{R}^4_{\theta}}$ of $\mathbb{R}^4_{\theta}$ by the action $\alpha$ of $\mathbb{R}^4$ is isomorphic to the $F^*$-crossed product $\mathcal{K}^{\infty}(L^2(\mathbb{C}^2)) \rtimes_{\widehat{\tau}} \mathbb{C}^2$, where $\widehat{\tau}$ is 
the dual action of $\tau$. Then it is isomorphic to $S(\mathbb{C}^2) \otimes 
\mathcal{K}^{\infty}(L^2(\mathbb{C}^2))$ as a $F^*$-algebra. \\
~~~~ We now consider a finitely generated projective right $\mathbb{R}^4_{\theta}$-module $\Xi$. Then there exist an integer $n \geq 1$ and a projection $P \in  M_n(\mathcal{M}(\mathbb{R}^4_{\theta}))$ such that $\Xi=P((\mathbb{R}^4_{\theta})^n)$. where $\mathcal{M}(\mathbb{R}^4_{\theta})$ is the $F^*$-algebra consisting of all bounded linear operators $T$ on $L^2(\mathbb{C}^2)$ whose kernel functions $T(\cdot~,~\cdot)$ are $\mathbb{C}$-valued bounded $C^{\infty}$-functions of $\mathbb{C}^2 \times \mathbb{C}^2$. 
Let us take the canonical faithful trace $Tr$ on $\mathbb{R}^4_{\theta}$ because of Proposition 1. Then we consider the moduli space: \\

\noindent
$\mathcal{M}^{(\mathcal{K}^{\infty}(L^2(\mathbb{C}^2)),\mathbb{C}^2,\alpha,Tr)}(\Xi)$ of $\Xi$ for $(\mathcal{K}^{\infty}(L^2(\mathbb{C}^2)),\mathbb{C}^2,\alpha,Tr)$. \\

We want to describe $P$ cited above as a precise fashion. Actually, we know that \[\mathcal{K}^{\infty}(L^2(\mathbb{C}^2)) \cong S(\mathbb{C}^2) \rtimes_{\lambda} \mathbb{C}^2 \]
\noindent
where $\cong$ means isomorphism as a $F^*$-algebra. $\lambda$ is the shift 
action of $\mathbb{C}^2$ on $S(\mathbb{C}^2)$. 
Then it follows that
\[ M_n(\mathcal{K}^{\infty}(L^2(\mathbb{C}^2))) \cong M_n(S(\mathbb{C}^2)) 
\rtimes_{\lambda^n} \mathbb{C}^2 \]
\noindent
where 
\[ \lambda^n_{w}(f)(w')=f(w'-w) \]
\noindent
for $f \in M_n(S(\mathbb{C}^2)),w,w' \in \mathbb{C}^2$. Let $\overline{\lambda^n}$ be the action of $\mathbb{C}^2$ on $M_n(\mathcal{K}^{\infty}(L^2(\mathbb{C}^2)))$ associated with $\lambda^n$ satisfying Theorem 2. 
It follows from  Proposition 3 that \\

$\mathcal{M}^{(\mathbb{R}^4_{\theta},\mathbb{R}^4,\alpha,Tr)}(\Xi) \approx \mathcal{M}^{(\mathcal{K}^{\infty}(L^2(\mathbb{C}^2),\mathbb{C}^2,\alpha,Tr)}(\Xi_1)$ \\

$~~~~~~~~~~~~~~~~~~~~ \approx \mathcal{M}^{(S(\mathbb{C}^2) \rtimes_{\lambda} \mathbb{C}^2,\mathbb{C}^2,\widehat{\lambda},\widehat{\int_{\mathbb{C}^2}dz})}(\Xi_2)~,$ \\
\noindent
where 
\[ \Xi_1=P_1(\mathcal{K}^{\infty}(L^2(\mathbb{C}^2)^n)~,~\Xi_2=P_2((S(\mathbb{C}^2) \rtimes_{\lambda} \mathbb{C}^2)^n) \]
\noindent
for the two projections $P_j~(j=1,2)$ with the property that 
\[ P_1 \in M_n(\mathcal{M}(\mathcal{K}^{\infty}(L^2(\mathbb{C}^2)))~,~P_2 \in 
M_n(\mathcal{M}(S(\mathbb{C}^2)) \rtimes_{\lambda} \mathbb{C}^2) \]
\noindent
corresponding to $\Xi$, where $\mathcal{M}(S(\mathbb{C}^2)$ is the $F^*$-algebra consisting of all $\mathbb{C}$-valued bouded $C^{\infty}$-functions on $\mathbb{C}^2$ and $\lambda$ is the shift action of $\mathbb{C}^2$ on $\mathcal{M}(S(\mathbb{C}^2))$. By its definition, we know that
\[\overline{\lambda}_w=\widehat{\lambda}_w \circ \widetilde{\lambda}_w ~,~(w \in \mathbb{C}^2) \]
\noindent
where $\widehat{\lambda}$ is the dual action of $\lambda$ and 
\[\widetilde{\lambda}_w(x)(w')=\lambda_wx(w') \]
\noindent
for all $x \in S(\mathbb{C}^2) \rtimes_{\lambda} \mathbb{C}^2$ and $w,w' \in \mathbb{C}^2$. Hence $\widehat{\lambda}$ commutes with $\overline{\lambda}$, which implies by Theorem 2 that there exist a $F^*$-dynamical system $(S(\mathbb{C}^2),\mathbb{C}^2,\widehat{\lambda}_{S(\mathbb{C}^2)},\int_{\mathbb{C}^2}dz)$ and a finitely generated projective right $A$-module $(\Xi_2)_{S(\mathbb{C}^2)}$ such that \\

\noindent
$ ~~~~~~~\mathcal{M}^{(S(\mathbb{C}^2) \rtimes_{\lambda} \mathbb{C}^2,\mathbb{C}^2,\widehat{\lambda},\widehat{\int_{\mathbb{C}^2}dz})}(\Xi_2)$ \\

$~~~~~~~~~~~\approx ~~~\mathcal{M}^{(S(\mathbb{C}^2),\mathbb{C}^2,\widehat{\lambda}_{S(\mathbb{C}^2)},\int_{\mathbb{C}^2}dz)}((\Xi_2)_{S(\mathbb{C}^2)}) ~.$ \\

\noindent
We know that there exist an integer $m \geq 1$ and a projection $Q \in M_m(\mathcal{M}(S(\mathbb{C}^2)))$ such that
\[ (\Xi_2)_{S(\mathbb{C}^2)}=Q(S(\mathbb{C}^2)^m)~.\]
\noindent
Moreover, it follows from the definition that the action 
$\widehat{\lambda}_{S(\mathbb{C}^2)}$ is nothing but $\lambda$. We now determine the moduli space 
\[\mathcal{M}^{(S(\mathbb{C}^2),\mathbb{C}^2,\lambda,\int_{\mathbb{C}^2}dz)})(Q(S(\mathbb{C}^2)^m)\] 
\noindent
in what follows: Since $Q \in M_m(\mathcal{M}(S(\mathbb{C}^2)))$ and 
\[ M_m(S(\mathbb{C}^2)) \cong S(\mathbb{C}^2,M_m(\mathbb{C})) ~,\]
\noindent
then it also follows from Theorem 2 that there exist a finitely generated projective right $\mathbb{C}$-module $Q(S(\mathbb{C}^2)^m)_{\mathbb{C}}$ such that \\

$~~~~~~~~\mathcal{M}^{(S(\mathbb{C}^2),\mathbb{C}^2,\widehat{\lambda}_{S(\mathbb{C}^2)},\int_{\mathbb{C}^2}dz)}(Q(S(\mathbb{C}^2)^m)) ~$  \\

$~~~~~~~~~~~\approx ~~~\mathcal{M}^{(\mathbb{C},\mathbb{C}^2,\lambda_{\mathbb{C}},1)}(Q(S(\mathbb{C}^2)^m)_{\mathbb{C}}) ~.$ \\

\noindent
Since $Q(S(\mathbb{C}^2)^m)_{\mathbb{C}}$ is a finitely generated projective right $\mathbb{C}$-module, then its construction tells us that there exists a projection $R \in M_m(\mathbb{C})$ such that 
\[  Q(S(\mathbb{C}^2)^m)_{\mathbb{C}})~=~R(\mathbb{C}^m) ~.\]
\noindent
Summing up the argument discussed above, we deduce that
\[ \mathcal{M}^{(\mathbb{R}^4_{\theta},\mathbb{R}^4,\alpha,Tr)}(\Xi) \approx \mathcal{M}^{(\mathbb{C},\mathbb{C}^2,\iota,1)}(R(\mathbb{C}^m))~.\]
\noindent
By the definition of the moduli space, we deduce that \\

$~~~~\mathcal{M}^{(\mathbb{C},\mathbb{C}^2,\iota,1)}(R(\mathbb{C}^m))$ \\

$~~~~~~~~~~~~~~\approx \mathrm{End}_{\mathbb{C}}(R(\mathbb{C}^m))_{sk}/U(\mathrm{End}_{\mathbb{C}}(R(\mathbb{C}^m))~,$ \\

\noindent
where 
\[\mathrm{End}_{\mathbb{C}}(R(\mathbb{C}^m))_{sk}~~~\mathrm{and}~~~U(\mathrm{End}_{\mathbb{C}}(R(\mathbb{C}^m)))\]
\noindent
are the set of all skew adjoint and unitary elements in \\
$\mathrm{End}_{\mathbb{C}}(R(\mathbb{C}^m))$ respectively.
Since $\mathrm{End}_{\mathbb{C}}(R(\mathbb{C}^m))=M_k(\mathbb{C})$ for some 
natural number $k~(m \geq k)$, it follows by using diagonalization that
\[\mathrm{End}_{\mathbb{C}}(R(\mathbb{C}^m))_{sk}/U(\mathrm{End}_{\mathbb{C}}(R(\mathbb{C}^m)) \approx \mathbb{R}^k ~,\]
\noindent
which implies the following theorem: \\

Theorem 2. \quad Let $\mathbb{R}^4_{\theta}$ be the deformation quantization of   $\mathbb{R}^4$ with respect to a skew symmetric matrix $\theta$ and take the  $F^*$-dynamical system $(\mathbb{R}^4_{\theta},\mathbb{R}^4,\alpha)$ with a 
canonical faithful $\alpha$-invariant trace $Tr$ of $\mathbb{R}^4_{\theta}$, 
where $\alpha$ is the translation action of $\mathbb{R}^4$ on $\mathbb{R}^4_{\theta}$. Suppose $\Xi$ is a finitely generated projective right $\mathbb{R}^4_{\theta}$-module, then there exists a natural number $k$ such that 
\[\mathcal{M}^{(\mathbb{R}^4_{\theta},\mathbb{R}^4,\alpha,Tr)}(\Xi) \approx 
\mathbb{R}^k  ~.\]

Remark. \quad The above theorem only states the topological data of the moduli 
spaces of Yang-Mills connections. We would study their both differential and 
holomorphic structures in a forthcoming paper (cf:[7]). \\

\pagebreak
\begin{center}
Acknowledgement
\end{center}

I would like to express my sincere gratitude to Professor T.Natsume for his 
careful reading and many pieces of advice to my manuscript, to Dr.S.Satomi
for his valuable comments, and to Ms.J.Takai for her constant encouragement.
\vspace{1cm}

\begin{center}
References
\end{center}

\vspace{1cm}
\noindent
[1]~A.Connes: An Analogue of the Thom Isomorphism for Crossed \\
~~~~Products of a C*-Algebra by an Action of $\mathbb{R}$,~Adv.Math.39,\\
~~~~(1981),31-55. \\

\noindent				  
[2]~A.Connes, M.R.Douglas and A.Schwarz: Noncommutative \\
~~~~Geometry and Matrix Theory: Compactification on Tori,\\
~~~~JHEP, 9802(1998)003.\\

\noindent				  
[3]~A.Connes and M.A.Rieffel:Yang-Mills for noncommutative \\
~~~~two tori,~Contemp.Math.Oper.Alg.Math.Phys.62,~A.M.S.\\
~~~~(1987),237-266.\\

\noindent
[4]~G.A.Elliott,T.Natsume and R.Nest: Cyclic cohomology for \\
~~~~one parameter smooth crossed products,Acta Math.,160 \\
~~~~(1988), 285-305. \\

\noindent
[5]~K.Furuuchi:~Instantons on Noncommutative $\mathbb{R}^4$ and \\
~~~~Projection Operators. ~arXiv:hep-th/9912047. \\

\noindent
[6]~H.Kawai: Constructive Formulation of String Theory,\\
~~~~Math.Sci.,4(2002),41-48. \\

\noindent
[7]~H.Nakajima:~Resolutions of Moduli Spaces of Ideal \\
~~~~Instantons on $\mathbb{R}^4$.~World Scientific.129-136(1994).\\

\noindent
[8]~N.Nekrasov and A.Schwarz:~Instantons on Noncom-\\
~~~~mutative $\mathbb{R}^4$,and $(2,0)$ Superconformal Six Dimen-\\
~~~~sional Theory.~Commun.Math.Phy.198,689-703(1998),\\

\noindent
[9]~Y.Ohkawa,~Matrix Models of M-Theory,Math.Sci.,\\
~~~~4(2002),35-40. \\

\noindent
[10]~N.Ohta, Duality of Superstring Theory and M-Theory,\\
~~~~~~Math.Sci.,4(2002),16-22. \\

\noindent
[11]~C.R.Putnam:~Commutation Properties of Hilbert Space \\
~~~~~~Operators and Related Topics.~Springer-Verlag (1967).\\

\noindent
[12]~M.A.Rieffel:~Deformation Quantization for Actions of \\
~~~~~~$\mathbb{R}^d$.~Memoires AMS.506(1993).\\

\noindent
[13]~H.Takai:~Yang-Mills Theory for Noncommutative Flows. \\
~~~~~~arXiv:math-ph/0403026.\\

\noindent
[14]~H.Takai:~Yang-Mills Theory for Noncommutative Flows, \\
~~~~~~Addendum.~arXiv:math-ph/0407038.

\end{document}